\documentclass[12pt]{article}
\usepackage[margin=1in]{geometry}
\usepackage{amsmath,amsthm,amssymb}
\usepackage{graphicx}
\usepackage{siunitx}
\usepackage[authoryear]{natbib}
\usepackage{booktabs}
\usepackage{multirow}
\usepackage{xcolor}
\usepackage{listings}
\usepackage{enumitem}
\usepackage{url}
\usepackage{cleveref}
\usepackage{tikz}
\usepackage{setspace}

\begin{document}

\title{A Critical Assessment of PINNs and Operator Learning for Geotechnical Engineering}
\author{Krishna Kumar}
\date{}
\maketitle
\begin{abstract}
Scientific machine learning (SciML) offers neural-network alternatives to numerical workflows in geotechnical engineering. This paper benchmarks multi-layer perceptrons (MLPs), physics-informed neural networks (PINNs), deep operator networks (DeepONet), and graph network simulators (GNS) against finite-difference and particle-based references on geotechnical benchmarks, and compares PINN inversion with automatic differentiation (AD) through a conventional solver. We evaluate each method for extrapolation, training and inference cost, transfer across problem instances, and physics accuracy. An MLP trained on two years of Terzaghi consolidation fits the data but at year ten predicts $\sim$290 mm with ReLU and $\sim$60 mm with tanh or sigmoid, against a reference of 99.3 mm. A PINN on a damped oscillator over $t \in [0,1]$ matches the closed form within that interval but fails outside, since the residual constrains the fit only where it is sampled. For the 1D wave equation, PINN training is $\sim$96{,}000 times slower than finite-difference methods and less accurate. DeepONet avoids PINN retraining, yet for the beam on elastic foundation its training cost equals $\sim$1.8 million finite-difference solves, and inference is slower per query than the direct solver. GNS improves geometric transfer through local particle interactions, though formulations still need trajectories, large training sets, and substantial memory. In the inverse wave benchmark, AD through the finite-difference solver recovers the material profile in seconds with $\sim$1\% error. The results support a cautious role for SciML. Neural networks suit interpolation and pattern recognition inside validated domains, while inverse analysis should first try differentiable physics-based solvers when a reliable forward solver exists.
\end{abstract}

\section{Introduction}
Scientific machine learning has become a common part of the geotechnical literature, with applications ranging from lateral spreading and CPT interpretation to regional site characterization and site response modeling \citep{Durante2021lateral, Hudson2023CPT, geyin2023us, ilhan2025artificial}. Within this broader literature, deep learning methods now appear in roles closer to numerical analysis than to pattern recognition, including PINNs for wave propagation and consolidation, DeepONet for foundation response, and GNS for granular flow. The central engineering question is therefore comparative. A trained network may evaluate quickly, but its value depends on the training cost, the accuracy of the resulting approximation, its behavior outside the calibration domain, and the amount of problem-specific retraining a new site or loading condition requires.

This paper evaluates those issues through direct numerical comparisons on canonical problems. The benchmarks are intentionally simple and include Terzaghi consolidation, a damped harmonic oscillator, the 1D wave equation, and a beam on elastic foundation. These problems have clear physics, exact or well-resolved numerical reference solutions, and direct relevance to settlement, wave propagation, and soil-structure interaction. Their low dimensionality also sharpens the assessment, because it removes much of the ambiguity that accompanies complex three-dimensional simulations. A method that is inaccurate, expensive, or unstable on these controlled tests needs additional evidence before engineers rely on it in settings where validation data are sparse and the physical state is only partially observed.

Recent reviews and position papers identify data requirements, physical consistency, extrapolation reliability, and scale as unresolved issues for machine learning in granular and geotechnical modeling \citep{Wang2025GranularMLreview, Fransen2026ScientificML}. The numerical experiments in this paper connect those issues to the function class defined by a neural-network architecture and to the soft, sampled nature of residual-based training. The assessment distinguishes tasks where machine learning is technically justified from tasks where a mature physics-based solver remains the better baseline. The comparison uses the same criteria that would apply to any engineering method, namely accuracy, computational cost, transfer across problem instances, physical consistency, and validation under spatially correlated data.

The organization follows the sequence of methods. Section~\ref{sec:mlp} uses MLPs to examine extrapolation and spatial validation. Section~\ref{sec:pinn} studies PINNs on a damped oscillator and the 1D wave equation. Section~\ref{sec:ad} introduces automatic differentiation and compares PINN inversion with AD through a finite difference solver. Section~\ref{sec:deeponet} evaluates DeepONet as a global operator between function spaces. Section~\ref{sec:gns} discusses GNS as a local, graph-based simulator for particle dynamics. Section~\ref{sec:discussion} synthesizes the results into guidance for geotechnical use.

\section{Multi-Layer Perceptrons in Geotechnical Engineering}
\label{sec:mlp}

\begin{figure}[h]
\centering
\includegraphics[width=0.85\textwidth]{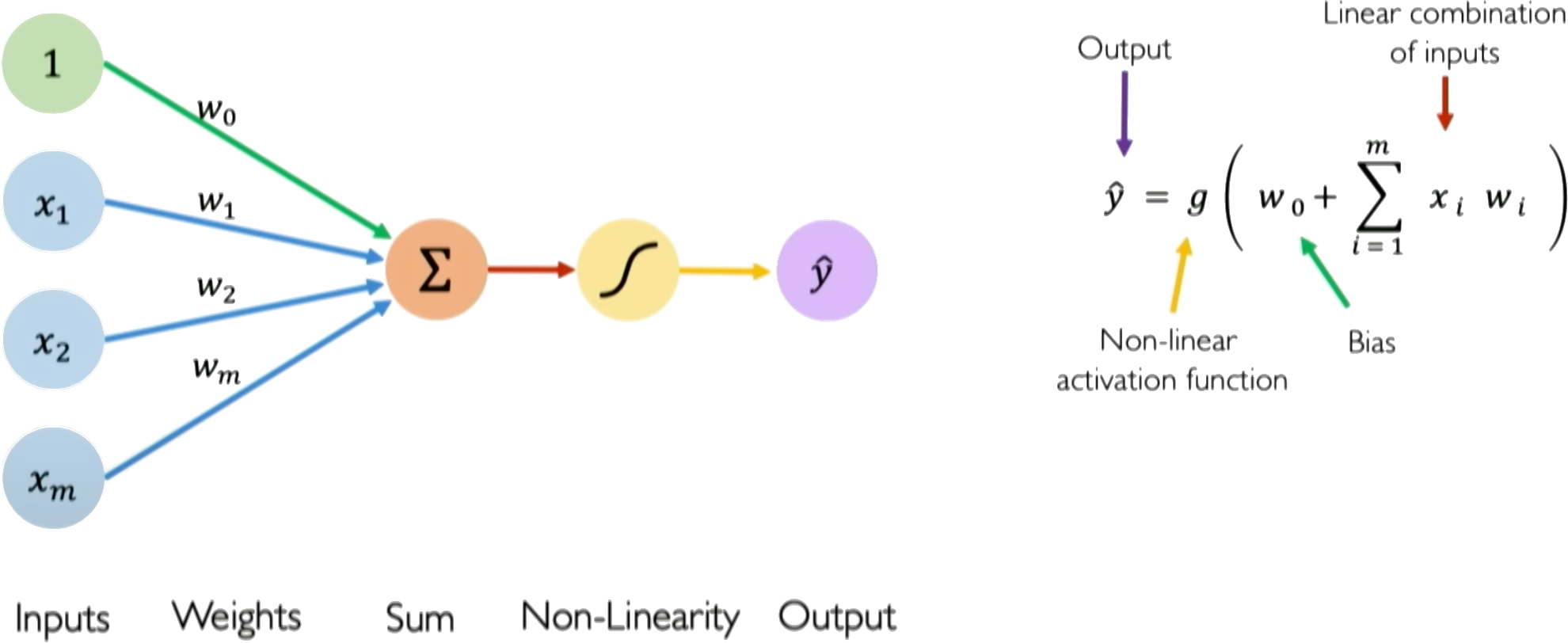}
\caption{A single neuron (perceptron) computes a weighted sum of inputs plus bias, then applies a nonlinear activation function $g$.
Stacking neurons into layers creates a multi-layer perceptron that can approximate complex nonlinear relationships.}
\label{fig:perceptron}
\end{figure}

A neural network represents a parameterized function. The simplest form we consider is the multi-layer perceptron (MLP), which also underlies the PINN, DeepONet, and GNS variants we discuss later. An MLP transforms inputs into outputs through layers of interconnected neurons, each of which computes a weighted sum of its inputs, adds a bias term, and passes the result through a nonlinear activation function $g$, as shown in \cref{fig:perceptron}.
\begin{equation}
\hat{y} = g\left(w_0 + \sum_{i=1}^{m} w_i x_i\right)
\end{equation}
Here $x_i$ are inputs such as depth, cone resistance, or pore pressure; $w_i$ are learnable weights; $w_0$ is the bias; and $g$ is the activation function.

Training changes the weights and biases, while the activation function $g$ is fixed before training and remains unchanged throughout. The network therefore learns by adjusting linear combinations of fixed nonlinear basis functions. For regression, the usual training objective is the mean squared error,
\begin{equation}
\mathcal{L} = \frac{1}{N}\sum_{i=1}^{N}(y_i - \hat{y}_i)^2
\end{equation}
Gradient descent updates the weights iteratively, $w \leftarrow w - \eta \nabla_w \mathcal{L}$, with $\eta$ the learning rate. Backpropagation computes those gradients by applying the chain rule layer by layer from output to input. Multiple layers enlarge the class of functions that can be represented, but they do not remove the dependence of extrapolation behavior on the chosen architecture and activation functions.

\subsection{Activation Functions}

\begin{figure}[htbp]
\centering
\includegraphics[width=0.75\textwidth]{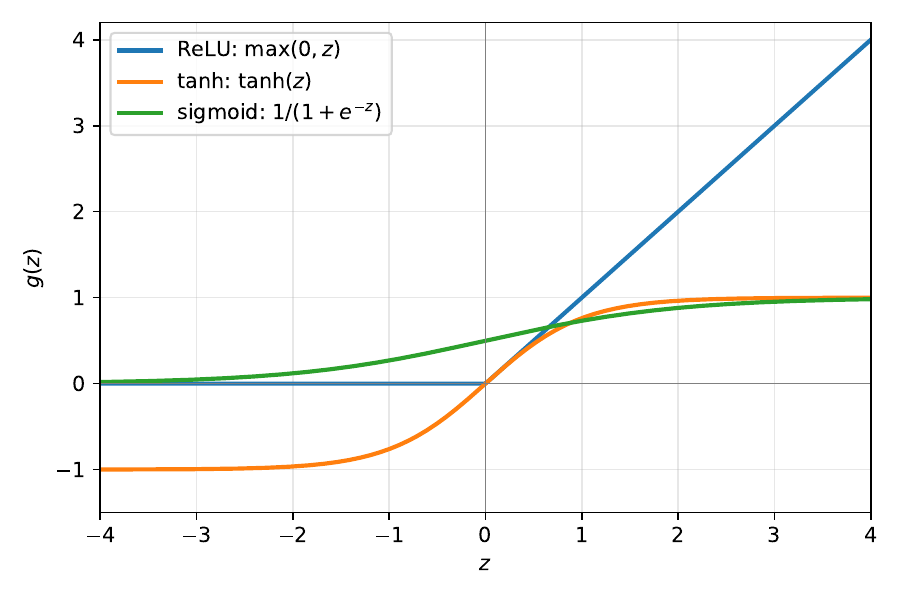}
\caption{The three activation functions used in the extrapolation experiment. ReLU grows linearly above zero, tanh saturates at $\pm 1$, and sigmoid saturates at 0 and 1. The tail shapes determine how a network extrapolates beyond its training range.}
\label{fig:activations}
\end{figure}

The activation function $g$ fixes the shapes from which the MLP constructs its approximation. Three common choices are shown in \cref{fig:activations}. ReLU, $g(z) = \max(0, z)$, is zero for negative inputs and grows linearly for positive inputs. Tanh, $g(z) = \tanh(z)$, saturates at $\pm 1$. Sigmoid, $g(z) = 1/(1 + e^{-z})$, saturates at 0 and 1. Inside the training range the optimizer combines neurons of these shapes to fit the observations. Outside that range the tails of the activation functions exert greater control, leading to linear growth for ReLU and plateauing behavior for tanh and sigmoid in the consolidation example below.

\subsection{The Extrapolation Problem}

Consider a mat foundation for a high-rise building for which settlement measurements from an earlier structure on the same site cover loads up to 200 kPa, while the new building imposes 350 kPa. The engineering task is extrapolation because predictions are required outside the range represented in the observations. The settlement follows $S(t) = S_\infty(1 - e^{-\alpha t})$, where $S_\infty$ is the final settlement and $\alpha$ controls the rate. The curve starts at zero, rises rapidly, and approaches an asymptote.

\begin{figure}[h]
\centering
\includegraphics[width=\textwidth]{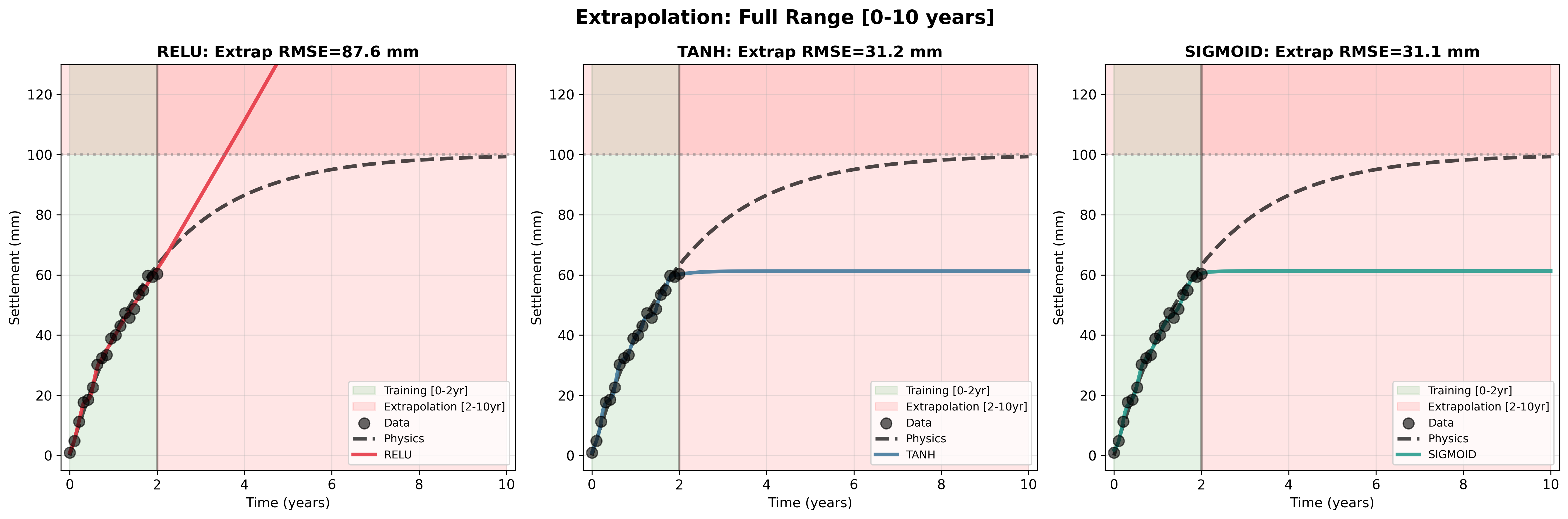}
\caption{Extrapolation failure for multi-layer perceptrons with different activation functions on consolidation settlement.
All three networks achieve good training accuracy on the 0 to 2 year training data (RMSE around 2 to 3 mm) but fail during extrapolation to 10 years.
ReLU produces unbounded linear growth (extrapolation RMSE = 104 mm), while tanh and sigmoid saturate prematurely (extrapolation RMSE = 32 mm).
The true solution asymptotically approaches 100 mm.}
\label{fig:extrapolation}
\end{figure}

The numerical experiment uses 20 settlement observations from the first two years ($t = 0$ to $t = 2$ years) as the training set, which represents typical settlement-plate measurements after loading. We train three MLPs with the same architecture, two hidden layers of 32 neurons, and three different activation functions, namely ReLU, tanh, and sigmoid. All three networks fit the training data closely, with training RMSE around 2 to 3 mm (\cref{fig:extrapolation}). We then extend the prediction horizon from year 2 to year 10. With $S_\infty = 100$ mm and $\alpha = 0.5$ year$^{-1}$, the reference settlement at year 10 is 99.3 mm.

The ReLU network predicts roughly 290 mm at year 10, an error of 190 mm, and extrapolates linearly upward without bound (RMSE = 104 mm over the extrapolation region from year 2 to year 10).
ReLU is piecewise linear, and outside the region where the network learned nonlinear combinations it can only produce linear growth.
The tanh and sigmoid networks fail in the opposite direction.
They predict 60 to 61 mm at year 10, underestimating by 39\% (RMSE = 32 mm), because both activation functions saturate and lead the networks to predict that settlement has nearly stopped even though consolidation is still progressing.

All three networks have low training error, while the fitted function class, rather than the consolidation equation, governs the extrapolated predictions. The universal approximation theorem \citep{Cybenko1989, Hornik1989} guarantees approximation on a compact domain, which in this experiment is the training envelope. It does not constrain the prediction beyond that domain. The networks learned functions that match an exponential curve between $t = 0$ and $t = 2$ years; outside that interval, the architecture provides no preference for Terzaghi's asymptotic decay over linear growth or premature saturation.

The same difficulty becomes more severe as the dimension of the input space increases. \citet{Balestriero2021} show that in high-dimensional spaces, new query points are likely to fall outside the convex hull of the training data. A liquefaction classifier with features such as peak ground acceleration $a_{\max}$, shear wave velocity $V_{s30}$, groundwater depth $d_w$, and distance to waterway $d_r$ may appear to interpolate in each individual variable while still extrapolating in the joint feature space. Adding soil layer depths, fines content, earthquake magnitude, and distance to fault increases that effect. For geotechnical applications, where engineers often evaluate loads, storms, earthquakes, and soil profiles that are absent from the calibration data, low training error is therefore insufficient evidence of predictive reliability.

\subsection{Validation Under Spatial Correlation}

Train-test splits report meaningful accuracy only when the test points are statistically independent of the training points, an assumption that fails for spatially correlated geotechnical data.

Consider 200 CPT soundings from five sites around a city.
We use them to train a classifier that identifies liquefaction-susceptible layers from tip resistance $q_c$, sleeve friction $f_s$, and pore pressure $u_2$.
Following standard machine learning practice, we shuffle the soundings randomly, train on 160 (80\%), and hold out 40 (20\%) for testing.
The classifier achieves 92\% test accuracy.

The 92\% accuracy does not estimate performance at a new site if the split ignores spatial structure. Soil properties are spatially correlated, and a CPT sounding at one location provides information about nearby ground conditions. Geostatistical models quantify this dependence through semivariograms with horizontal correlation lengths commonly on the order of 50 to 100 meters \citep{Vanmarcke1977, Fenton2007}. Measurements separated by 10 meters may contain much of the same geological information, while measurements separated by larger distances or by different depositional environments may not.

\begin{figure}[h]
\centering
\includegraphics[width=\textwidth]{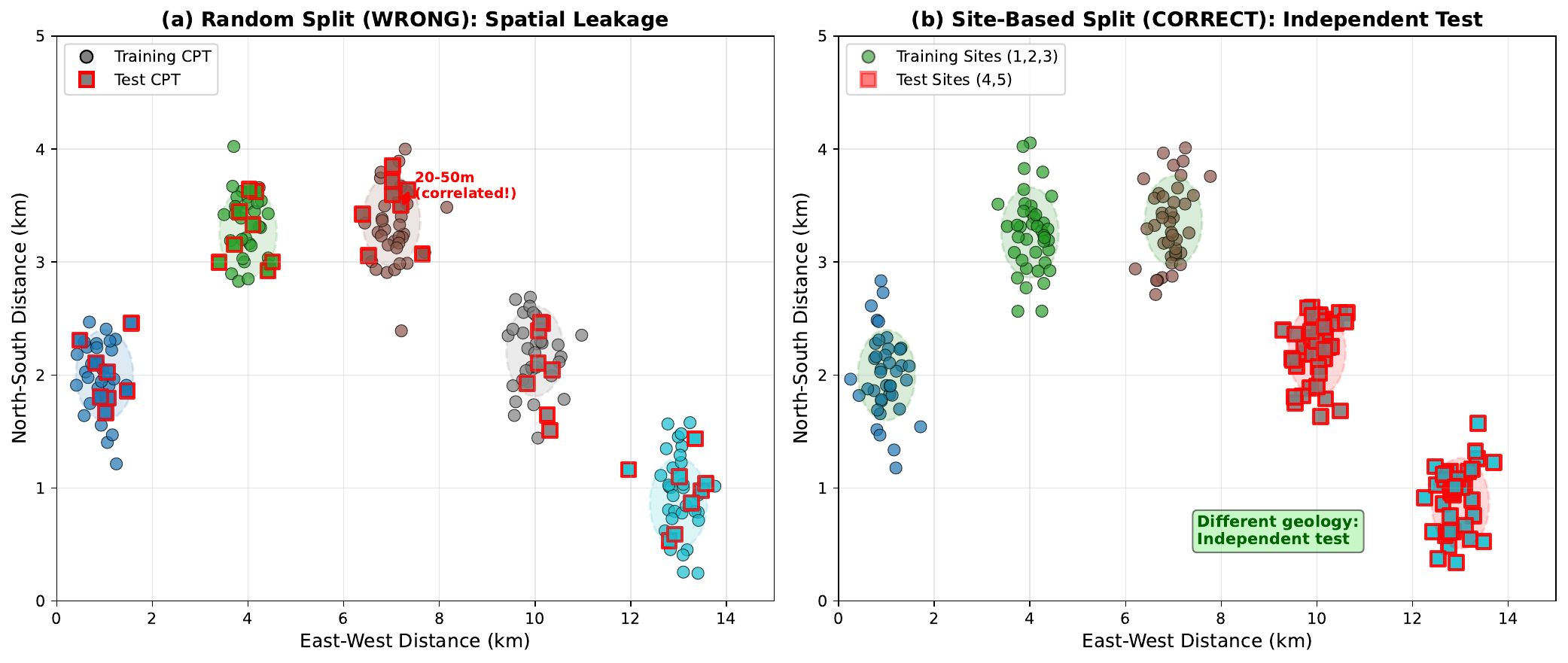}
\caption{Spatial autocorrelation in geotechnical data. (a) A random split creates spatial leakage because test points (red squares) lie within correlation ellipses of training points (circles), which inflates reported accuracy. (b) A site-based split ensures that test sites (4, 5) are geologically independent of training sites (1, 2, 3), which provides a more appropriate estimate of transfer to new sites.
Correlation lengths in soil properties are typically 50 to 100 m horizontally.}
\label{fig:spatial}
\end{figure}

A random split of 200 CPT soundings from five sites, with 40 soundings per site, gives each site roughly 32 training soundings and 8 test soundings. The test soundings from Site 1 may lie 20 to 50 meters from training soundings at the same site and can therefore fall inside the same spatial correlation ellipse (\cref{fig:spatial}a). In that case, the test set measures interpolation within known geological settings more than transfer to a new site.

Site-based splitting changes the validation question (\cref{fig:spatial}b). If Sites 1, 2, and 3 provide the 120 training soundings and Sites 4 and 5 provide the 80 test soundings, the test data differ in stratigraphy, depositional history, and groundwater conditions. Leave-one-site-out cross-validation applies the same principle when the number of sites is limited. This validation design estimates transfer to a new project location rather than interpolation among correlated measurements from sites already represented in training.

Geotechnical practice already contains tools for handling spatial dependence. Kriging \citep{Matheron1963} interpolates soil properties between boreholes while accounting for correlation structure, and random field models \citep{Fenton2007} embed correlation lengths and anisotropy into finite element analysis. Machine-learning workflows need the same geological discipline in their train-test design; otherwise, reported accuracy can reflect spatial leakage rather than deployable generalization.

Class imbalance creates a related validation issue.
High-consequence events are rare, and a typical liquefaction database might contain 200 liquefaction cases (4\%) alongside 4{,}800 non-liquefaction cases (96\%).
A classifier that predicts ``no liquefaction'' for every site achieves 96\% accuracy while missing every actual failure.
Models for such problems should report recall (the fraction of actual failures detected) and F1-score rather than accuracy.
In geotechnical applications, a missed failure (false negative) carries a much higher cost than a false alarm (false positive).

\section{Physics-Informed Neural Networks and Their Computational Cost}
\label{sec:pinn}

A physics-informed neural network (PINN) uses the same MLP function approximator discussed in \cref{sec:mlp}, but adds loss terms that penalize violation of a governing differential equation and its boundary or initial conditions. Since the introduction of PINNs by \citet{Raissi2019PINN}, geotechnical applications have included one-dimensional consolidation \citep{Bekele2021PINNconsolidation}, pile-soil interaction \citep{Ouyang2024PINNpile}, seismic site response \citep{Liu2025PINNseismic}, and broader geoengineering analysis \citep{Chen2024PINNgeoeng}.

For a differential equation $\mathcal{L} u = 0$ on a domain $\Omega$ with boundary or initial conditions $\mathcal{B} u = g$ on $\partial \Omega$, the network $f_\theta(x)$ maps coordinates directly to the predicted field value $u(x)$. Because $f_\theta$ is differentiable with respect to its inputs, automatic differentiation can evaluate the derivatives that appear in $\mathcal{L}$ (\cref{sec:ad}). Substitution of $f_\theta$ into the differential operator gives a pointwise residual $\mathcal{R}(x;\theta) = \mathcal{L} f_\theta(x;\theta)$, and training minimizes a composite loss,
\begin{equation}
\mathcal{L}_{\text{total}}(\theta) = \underbrace{\frac{1}{N}\sum_{i=1}^N |f_\theta(x_i) - u_i|^2}_{\mathcal{L}_{\text{data}}} + \lambda\, \underbrace{\frac{1}{N_c}\sum_{j=1}^{N_c} |\mathcal{R}(x_j^c;\theta)|^2}_{\mathcal{L}_{\text{phys}}} + \lambda_b\, \mathcal{L}_{\text{BC}}(\theta).
\label{eq:pinn_loss}
\end{equation}
The data loss matches $N$ measurements at points $x_i$.
The physics loss penalizes violation of the PDE at $N_c$ \emph{collocation points} $x_j^c$ sampled throughout $\Omega$.
The boundary loss enforces $\mathcal{B} u = g$ at boundary samples.
The weights $\lambda$ and $\lambda_b$ control the relative importance of each term.

The data term matches measurements at $x_i$, the physics term penalizes the PDE residual at collocation points $x_j^c$, and the boundary term enforces $\mathcal{B} u = g$ at sampled boundary points. These residual and boundary terms add information between sparse measurements, but they remain sampled penalties on an MLP function class. The loss therefore constrains the trained network most directly at the points and regions where it samples the equation. \Cref{sec:oscillator_test} uses a damped harmonic oscillator to examine how that distinction affects extrapolation beyond the collocation interval.

A trained PINN also serves only the problem instance it learned. Any change to the initial condition, boundary location, or material profile requires retraining, because the learned weights approximate one solution field rather than a reusable numerical procedure. A finite difference solver handles those changes through input arrays and boundary-condition data, while a PINN repeats a nonlinear optimization for each new forward problem.

The residual also enters as a soft constraint. Training must balance PDE, boundary, and data losses whose magnitudes can differ substantially, and neural networks tend to learn smooth components before sharp gradients, a behavior known as spectral bias. These effects produce multi-objective imbalance during PINN training \citep{Krishnapriyan2021, Wang2021PINN}. The wave-equation experiment in \cref{sec:wave_cost} measures the computational cost of that training problem.

\subsection{Testing the Physics Claim on a Damped Harmonic Oscillator}
\label{sec:oscillator_test}

\begin{figure}[htbp]
\centering
\begin{minipage}{0.48\textwidth}
\centering
\textbf{(a)} Data-only fit\\
\includegraphics[width=\textwidth]{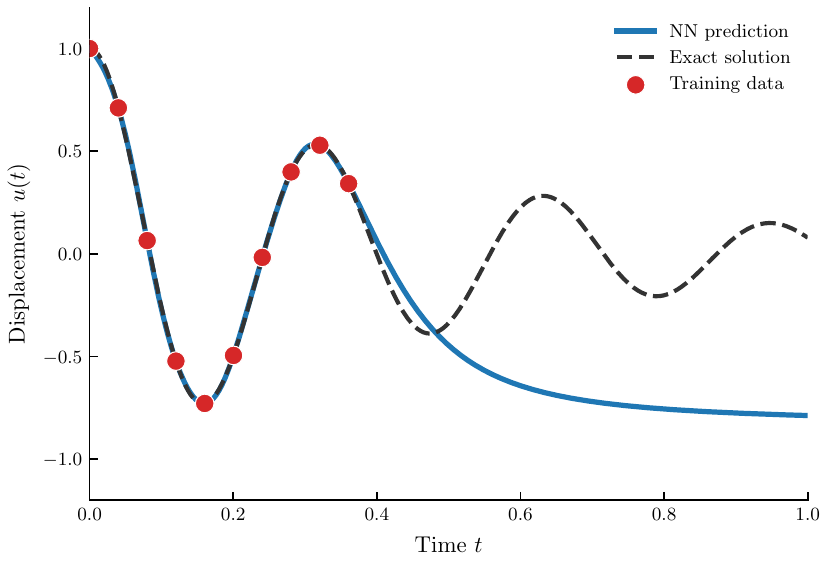}
\end{minipage}
\hfill
\begin{minipage}{0.48\textwidth}
\centering
\textbf{(b)} PINN with physics loss\\
\includegraphics[width=\textwidth]{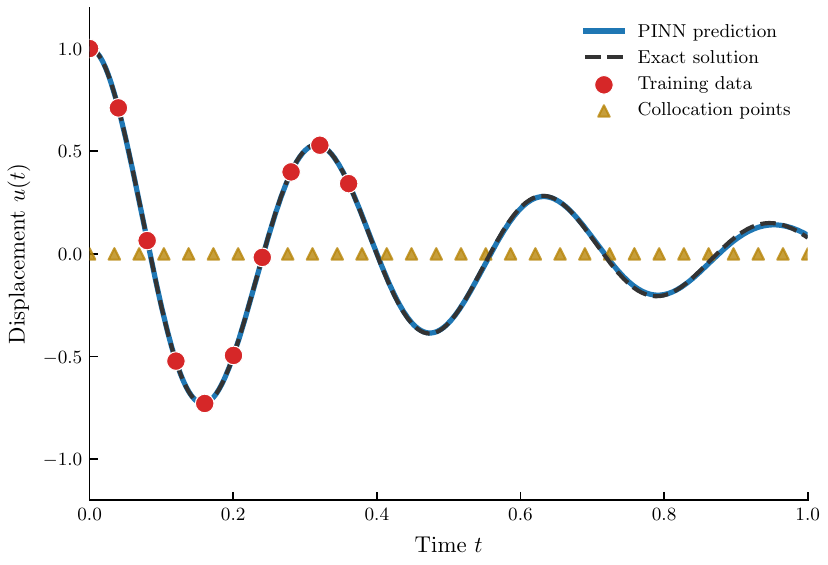}
\end{minipage}
\caption{Inside the training interval $[0,1]$. (a) A neural network trained only on the ten sparse measurements (red) fits the data but diverges between them. (b) Adding the ODE residual at collocation points recovers the closed-form solution (dashed grey).}
\label{fig:oscillator_inside}
\end{figure}

\begin{figure}[htbp]
\centering
\includegraphics[width=0.7\textwidth]{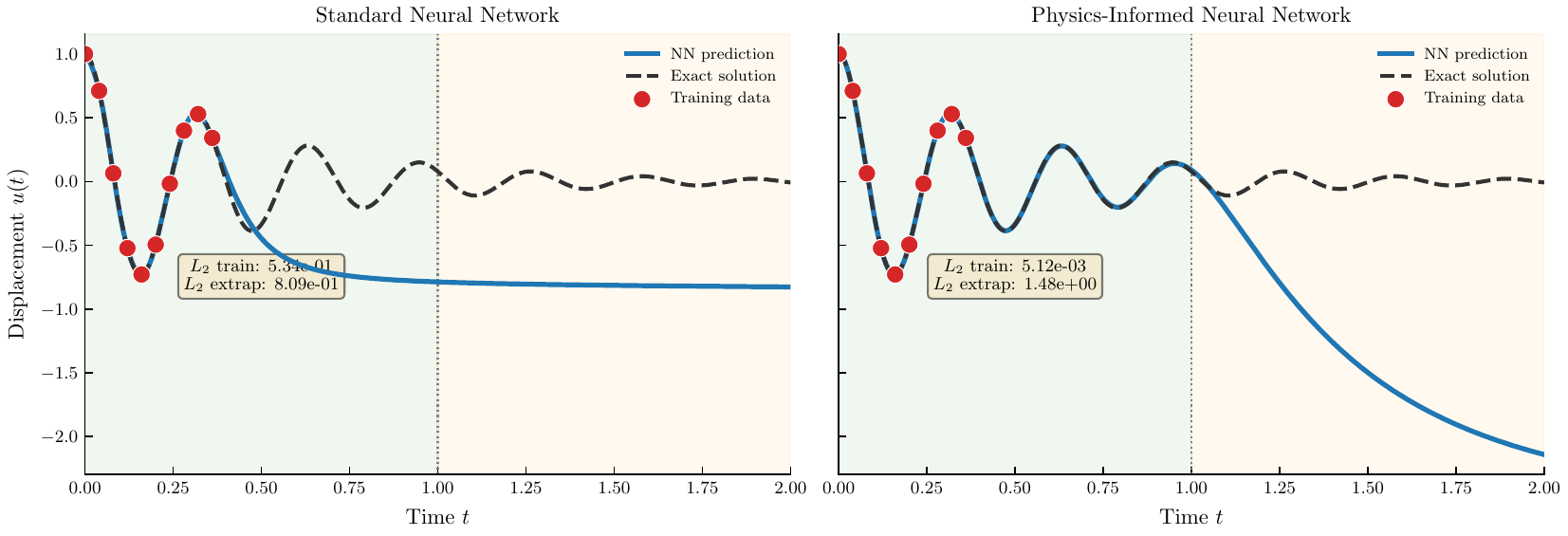}
\caption{PINN trained on $t \in [0,1]$ and evaluated on $t \in [0,2]$. Inside the training interval the prediction matches the exact solution. Outside, where no collocation points enforce the ODE, the prediction drifts and the oscillation is lost.}
\label{fig:oscillator_extrap}
\end{figure}

The first PINN experiment uses a damped harmonic oscillator,
\begin{equation}
\ddot{u} + 2\delta\dot{u} + \omega_0^2 u = 0, \qquad u(0) = 1, \quad \dot{u}(0) = 0,
\label{eq:oscillator}
\end{equation}
with $\delta = 2$ s$^{-1}$ and $\omega_0 = 20$ rad/s.
The system is underdamped and oscillates at $\omega_d = \sqrt{\omega_0^2 - \delta^2} \approx 19.9$ rad/s while the amplitude decays exponentially.
Ten sparse measurements of $u(t_i)$ on $t \in [0, 0.36]$ form the training data, and 30 collocation points cover $t \in [0, 1]$.
We first train an MLP with three hidden layers of 32 neurons and tanh activations on the data loss alone. It fits the ten measurements but diverges between them (\cref{fig:oscillator_inside}a). Adding the ODE residual at the collocation points produces a PINN that reconstructs the closed-form solution across $t \in [0, 1]$ (\cref{fig:oscillator_inside}b). When we evaluate the same PINN on $t \in [0, 2]$, the prediction remains accurate over the collocation interval, drifts after $t = 1$, and loses the oscillatory solution by $t = 2$ (\cref{fig:oscillator_extrap}).

This extrapolation test clarifies the role of the physics loss. The residual term is a mean squared error evaluated at a finite set of sampled points, with targets generated from the differential equation. The collocation points in $[0,1]$ supply those residual targets, and the network learns a function that satisfies them over the represented interval. Outside that interval the architecture is still the MLP function class from \cref{sec:mlp}, without the recurrence or state update that a time-marching solver would apply. Agreement inside the collocation domain therefore cannot by itself establish that a PINN will respect the governing equation under extrapolation.

\subsection{Cost on the 1D Wave Equation}
\label{sec:wave_cost}

The oscillator test isolates extrapolation, and the next experiment measures cost and accuracy on the 1D wave equation $u_{tt} = c^2 u_{xx}$, which governs SH-wave propagation in soil and underlies seismic site response analysis.
We solve a normalized benchmark on $x \in [0,1]$ with constant wave speed $c = 0.95$.
The value is nondimensional and is not intended to represent a field-scale shear-wave velocity.
The initial condition is a Gaussian pulse $u(x,0) = \exp(-25(x-0.5)^2)$ at the center, with zero displacement at both boundaries.
The setup represents a pulse-propagation test problem rather than a calibrated site-response model.

\begin{figure}[h]
\centering
\resizebox{0.85\textwidth}{!}{\begin{tikzpicture}[
    input/.style={circle, draw, fill=blue!30, minimum size=50pt, inner sep=0pt},
    hidden/.style={circle, draw, fill=gray!50, minimum size=50pt, inner sep=0pt},
    output/.style={circle, draw, fill=green!30, minimum size=60pt, inner sep=0pt},
    deriv/.style={circle, draw, fill=yellow!30, minimum size=45pt, inner sep=0pt},
    box/.style={rectangle, draw, dashed, rounded corners, minimum width=80pt, minimum height=240pt},
    pde/.style={rectangle, draw, fill=orange!20, minimum width=180pt, minimum height=50pt, rounded corners, align=center},
    conn/.style={gray!50, line width=1pt},
    arrow/.style={->, >=latex, thick}
]
    \node[text width=3cm, align=center] at (0,6.5) {Input Layer\\(Space-Time)};
    \node[text width=3cm, align=center] at (4,6.5) {Hidden Layer 1};
    \node[text width=3cm, align=center] at (8,6.5) {Hidden Layer 2};
    \node[text width=3cm, align=center] at (12,6.5) {Output Layer\\Displacement};

    \node[input] (x) at (0,1) {\Huge $x$};
    \node[input] (t) at (0,-1) {\Huge $t$};

    \foreach \y [count=\i] in {4,2,0,-2,-4}
        \node[hidden] (h1-\i) at (4,\y) {\large $z_{\i}^1$};

    \foreach \y [count=\i] in {4,2,0,-2,-4}
        \node[hidden] (h2-\i) at (8,\y) {\large $z_{\i}^2$};

    \node[output] (u) at (12,0) {\Large $u(x,t)$};

    \node[deriv] (u_out) at (16,3) {\large $u$};
    \node[deriv] (dudt) at (16,1.5) {\large $\frac{\partial u}{\partial t}$};
    \node[deriv] (d2udt2) at (16,0) {\large $\frac{\partial^2 u}{\partial t^2}$};
    \node[deriv] (dudx) at (16,-1.5) {\large $\frac{\partial u}{\partial x}$};
    \node[deriv] (d2udx2) at (16,-3) {\large $\frac{\partial^2 u}{\partial x^2}$};

    \node[box] (deriv_box) at (16,0) {};

    \node[pde] (pde) at (15.5,-6) {Wave Equation: $\frac{\partial^2 u}{\partial t^2} = c^2 \frac{\partial^2 u}{\partial x^2}$\\Compute derivatives using AD\\ and minimize PDE residual + BC + IC};

    \foreach \i in {1,...,5} {
        \draw[conn] (x) -- (h1-\i);
        \draw[conn] (t) -- (h1-\i);
    }

    \foreach \i in {1,...,5}
        \foreach \j in {1,...,5}
            \draw[conn] (h1-\i) -- (h2-\j);

    \foreach \i in {1,...,5}
        \draw[conn] (h2-\i) -- (u);

    \draw[arrow] (u) -- (u_out);
    \draw[arrow] (u) -- (dudt);
    \draw[arrow] (u) -- (d2udt2);
    \draw[arrow] (u) -- (dudx);
    \draw[arrow] (u) -- (d2udx2);

    \draw[arrow] (deriv_box) -- (pde);
\end{tikzpicture}}
\caption{PINN architecture for the 1D wave equation.
The network takes spatial $(x)$ and temporal $(t)$ coordinates as inputs, passes them through hidden layers, and outputs displacement $u(x,t)$.
Automatic differentiation computes the spatial and temporal derivatives ($\partial u/\partial x$, $\partial^2 u/\partial x^2$, $\partial u/\partial t$, $\partial^2 u/\partial t^2$) that feed the wave equation residual. The loss combines the PDE residual with initial and boundary condition terms.}
\label{fig:pinn_arch}
\end{figure}

The forward wave benchmark compares a standard finite difference (FD) solver with a PINN. The FD solver uses 1000 spatial grid points and 5000 time steps with explicit central differences, the approach behind most geotechnical seismic codes.
The PINN has 4 hidden layers of 128 neurons each.
We use sine activations (SIREN with frequency parameter $\omega_0 = 30$) to capture the high-frequency oscillations that ReLU and tanh cannot.
The network takes $(x, t)$ normalized to $[-1, 1]$ and outputs the scalar field $u(x,t)$.
We train the PINN for 9000 Adam iterations at learning rate $5 \times 10^{-4}$.
The training requires careful tuning of the learning rate schedule and loss weighting coefficients.

\begin{figure}[h]
\centering
\includegraphics[width=\textwidth]{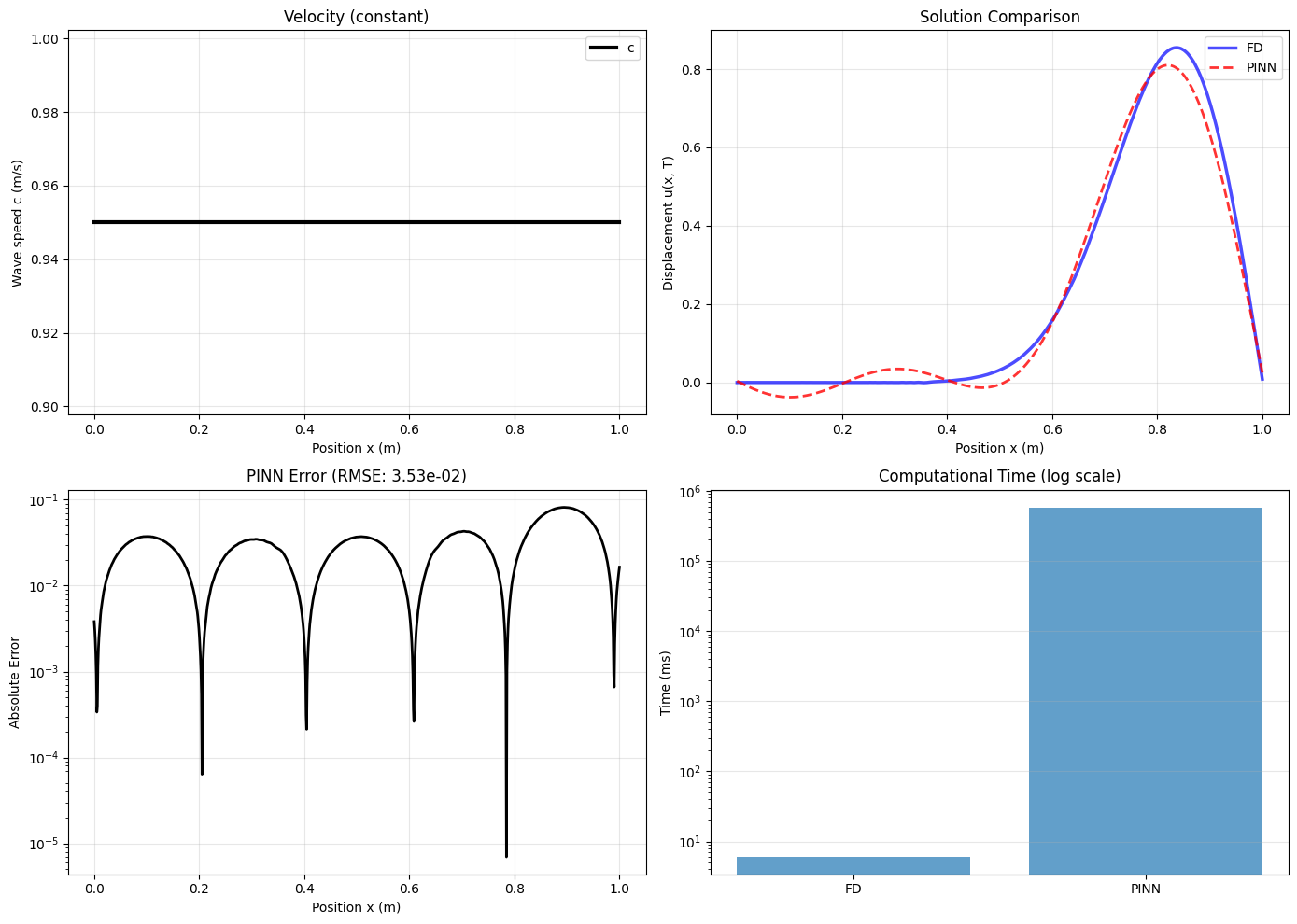}
\caption{Wave propagation comparison between PINN and finite difference.
The four-panel comparison shows (a) the constant normalized velocity profile $c = 0.95$, (b) the solution at final time $t = 2.5$ for FD (blue) versus PINN (red dashed), (c) pointwise error on a logarithmic scale showing PINN errors of order $10^{-2}$, and (d) computational time, in which the PINN is 96{,}372$\times$ slower than FD (6.04 ms versus 581.92 s) while achieving RMSE = $3.53 \times 10^{-2}$ against FD's machine precision.}
\label{fig:waveform}
\end{figure}

The finite difference solver completes in 6.04 milliseconds.
The PINN training takes 582 seconds, or almost ten minutes, which is about 96{,}000 times the finite difference runtime. At final time $t = 2.5$, the root-mean-square difference between PINN and FD solutions is $3.5 \times 10^{-2}$. The finite difference solution satisfies the discrete wave equation to machine precision (error $< 10^{-15}$), while the PINN residual remains around $10^{-3}$.
\cref{fig:waveform} shows the waveform comparison and error distribution.

The quantitative performance is summarized in \cref{tab:pinn}.

\begin{table}[h]
\centering
\caption{PINN performance for forward and inverse problems (1D wave equation).}
\label{tab:pinn}
\small
\begin{tabular}{llrrr}
\toprule
Problem type & Method & Time (s) & RMSE & Speedup \\
\midrule
\multirow{2}{*}{Forward} & Finite difference & 0.00604 & $<10^{-15}$ & 1$\times$ \\
                         & PINN & 582 & 0.035 & 96{,}000$\times$ slower \\
\midrule
\multirow{2}{*}{Inverse} & AD & seconds & order $10^{-3}$ & 1$\times$ \\
                         & PINN & comparable & order $10^{-2}$ & Similar, larger error \\
\bottomrule
\end{tabular}
\end{table}

The long training time and residual error are consistent with the competing objectives in the loss function. The optimizer minimizes PDE residuals, initial-condition errors, and boundary-condition errors simultaneously, and these terms have different magnitudes during training. In this experiment, the PDE residual drops quickly to $10^{-2}$, the boundary-condition term stagnates near $10^{-3}$, and the initial-condition error remains near $10^{-4}$. \citet{Wang2022PINN} describe this behavior as multi-objective imbalance, where training may reduce one part of the loss while leaving another physically important constraint under-satisfied.

Adaptive loss weighting, curriculum learning, and separate networks for different physics components can reduce some PINN training pathologies, but they also add algorithmic choices that must be validated for each problem. The finite difference comparison is therefore not only a timing baseline; it also represents a discretization with known consistency, stability, and convergence properties. In this benchmark, the FD implementation is about 80 lines, while the PINN implementation exceeds 500 lines and requires extensive hyperparameter tuning.

For forward problems on regular domains with known parameters, the comparison favors finite difference or finite element discretizations whenever those solvers already run in seconds. Under such conditions, a 96{,}000$\times$ training slowdown would require a compensating gain in transferability or accuracy that is absent from this benchmark.

\section{Automatic Differentiation and Differentiable Programming}
\label{sec:ad}

Automatic differentiation (AD) turns a traditional solver into a differentiable program and provides a useful baseline for inverse problems. AD computes derivatives of functions specified by computer programs by applying the chain rule to the elementary operations executed during the forward calculation. Numerical differentiation estimates sensitivities from finite perturbations, and symbolic differentiation manipulates mathematical expressions; AD instead differentiates the executed program. The resulting derivatives are exact up to machine precision for the implemented computation.

\begin{figure}[h]
\centering
\begin{minipage}{0.48\textwidth}
\centering
\includegraphics[width=\textwidth]{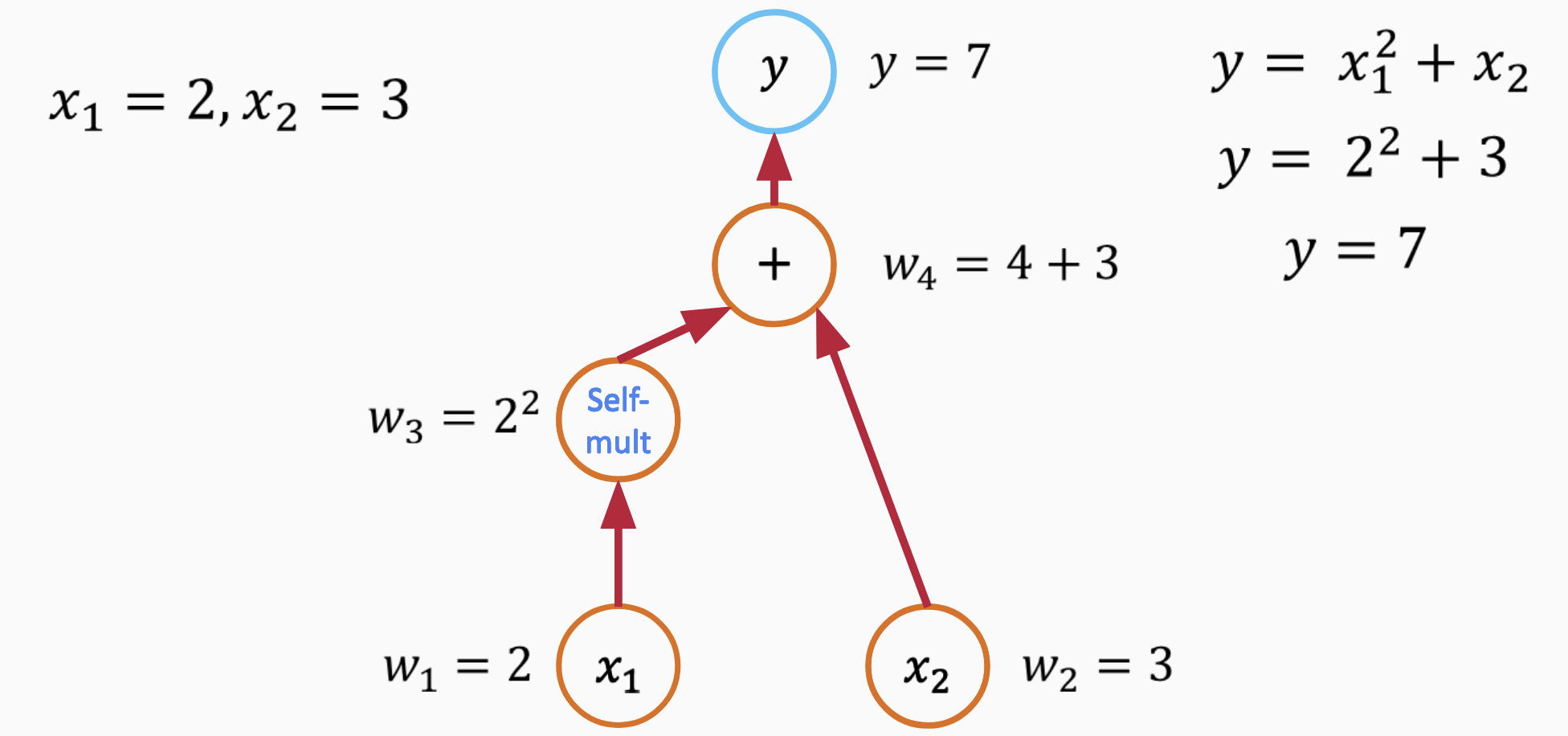}
\end{minipage}
\hfill
\begin{minipage}{0.48\textwidth}
\centering
\includegraphics[width=\textwidth]{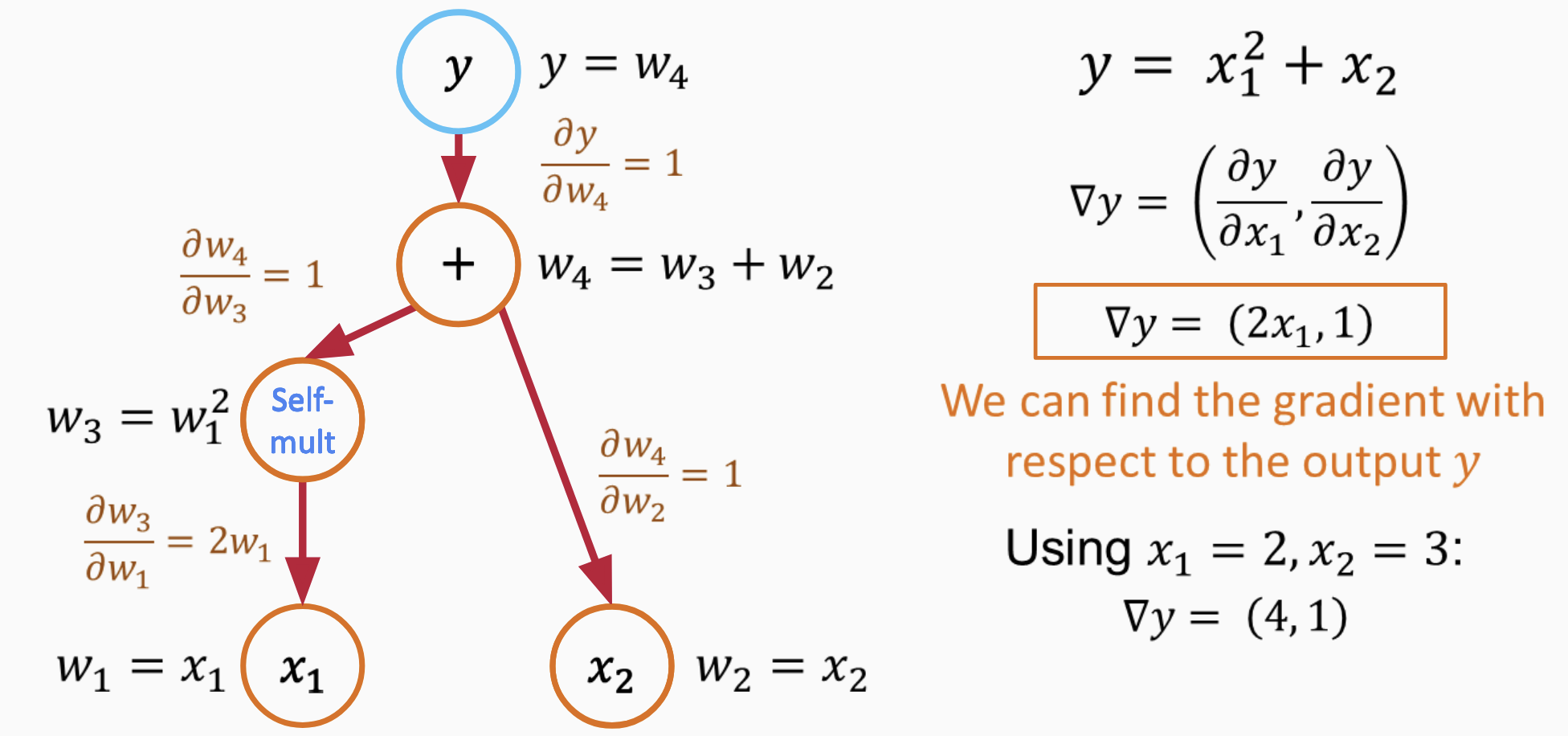}
\end{minipage}
\caption{Automatic differentiation modes.
Forward mode (left) accumulates derivatives alongside the forward computation.
Reverse mode (right) accumulates adjoints by traversing the computational graph backward.
For inverse problems with many parameters (inputs) and a single loss (output), reverse mode is far more efficient because it requires only one backward pass regardless of parameter count.}
\label{fig:ad_modes}
\end{figure}

AD operates in two modes, forward and reverse (\cref{fig:ad_modes}).
Forward mode computes derivatives in the same direction as the program execution and efficiently computes gradients when outputs outnumber inputs.
Reverse mode, called backpropagation in neural network contexts, traverses the computation graph backward and efficiently handles cases where inputs vastly outnumber outputs.
The reverse case applies to most inverse problems, which have many parameters but few objective function outputs.

In neural-network training, reverse-mode AD is more commonly called backpropagation. It computes the gradient of the loss with respect to every network weight in one backward pass through the layers, regardless of how many weights the network has, and gradient descent then updates the weights. The same machinery applied to a traditional solver computes the gradient of a loss with respect to physical parameters such as material moduli, a velocity profile, or boundary tractions. Inverse parameter identification then becomes a gradient-based search over those parameters, with derivatives computed exactly through the solver's discretized physics. We label this practice \emph{differentiable programming}: a numerical solver written so that AD can differentiate it end to end, producing outputs alongside exact gradients with respect to its inputs \citep{WangKumar2024LBM, Cheng2025MPM}.

We wrap finite difference codes for consolidation, wave propagation, or seepage in JAX or PyTorch and obtain gradient-based optimization for parameter estimation and inverse problems. The solver becomes an oracle that returns both the solution and exact sensitivities for any input. Differentiable programming therefore extends beyond machine learning and brings gradient-based optimization to traditional computational physics.

Differentiable programming has costs because reverse-mode AD stores every intermediate state of the forward computation, so memory grows with the length of the simulation.
A time-stepping solver with $10^5$ steps and $10^4$ degrees of freedom can require tens of gigabytes of activation storage.
Checkpointing trades memory for recomputation by saving state only at sparse checkpoints and recomputing intermediate steps on demand.
The implementation also demands that every operation in the solver is differentiable, which sometimes requires rewriting code that uses non-differentiable branching or external libraries.
AD is attractive for inverse problems because it differentiates the same solver used for the forward calculation, although its memory cost grows with simulation length and the implementation requires every solver operation to be differentiable.

\subsection{Inverse Problems with AD and PINNs}

The inverse problem uses the same wave equation, but now we recover an unknown normalized velocity profile $c(x)$ from the final wavefield $u(x, t_f)$.
The true profile rises linearly from $0.9$ at $x = 0$ to $1.0$ at $x = 1$.
The forward solver records $u(x, t_f)$ on the same 1000-point spatial grid that discretizes $c(x)$.

\noindent \paragraph{Automatic differentiation approach:} We wrap the finite difference solver in JAX and minimize the wavefield misfit $\mathcal{L}(c) = \|u_{\text{sim}}(x; c) - u_{\text{obs}}(x)\|^2$ directly over the 1000-point velocity vector $c$.
Starting from $c(x) = \text{linspace}(0.85, 1.0)$, we run 500 Adam iterations at learning rate $10^{-3}$ followed by 400 L-BFGS iterations.
Each iteration runs the forward solver once and backpropagates exact gradients through the entire computation, without approximation, soft constraints, neural network training, or multi-objective balancing.
The solver enforces physics exactly at every step, subject only to discretization error.

\paragraph{PINN inverse approach:} The inverse PINN imposes a linear parametric ansatz $c(x) = c_{\text{left}} + (c_{\text{right}} - c_{\text{left}})\, x$ and learns $c_{\text{left}}$ and $c_{\text{right}}$ alongside the network weights.
It simultaneously tries to match the observed wavefield, satisfy the PDE residual at collocation points, and recover the velocity parameters.
This formulation couples parameter estimation to the same multi-objective training problem that appears in the forward PINN.

\begin{figure}[htbp]
\centering
\includegraphics[width=0.95\textwidth]{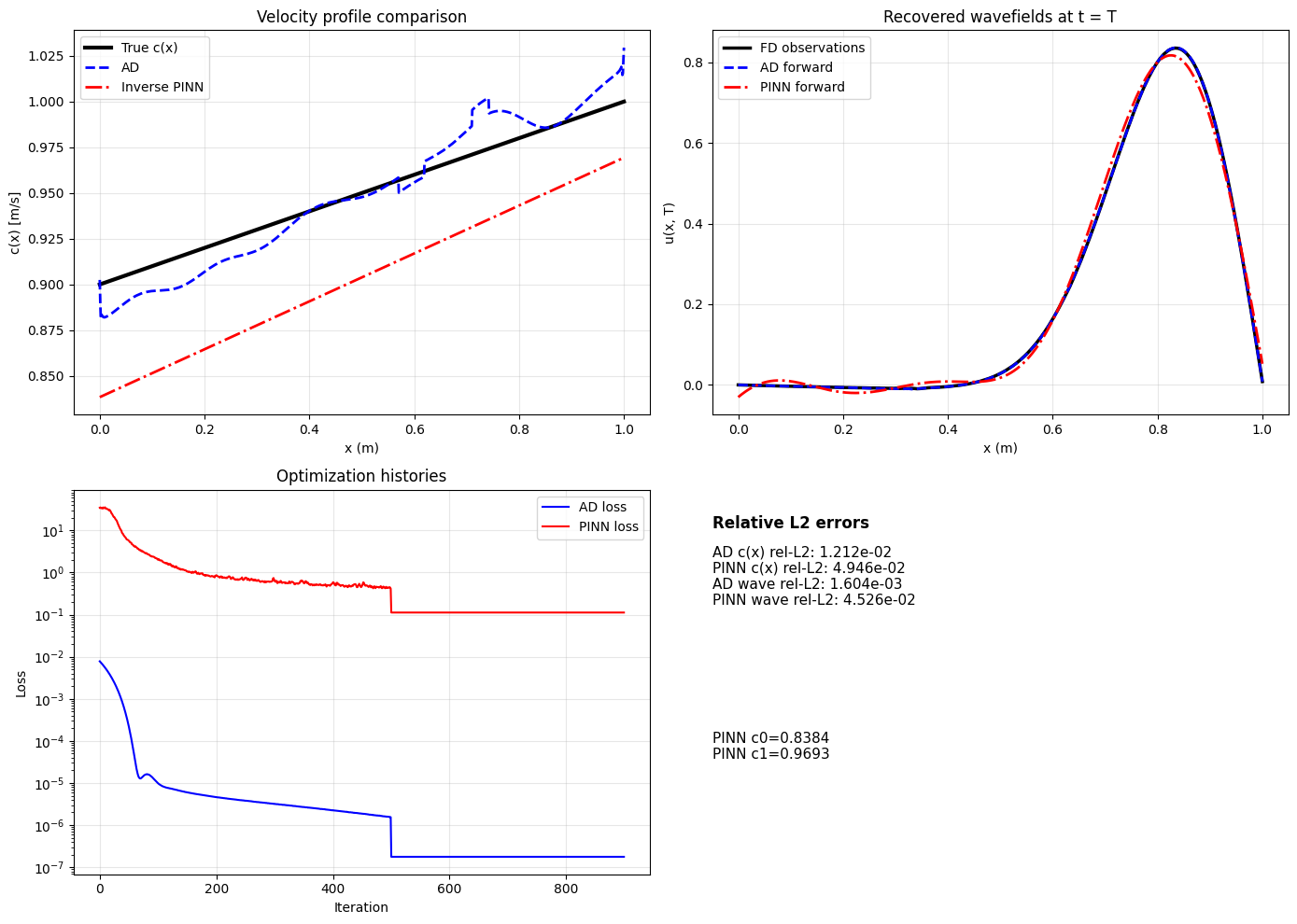}
\caption{Inverse problem comparison for recovering velocity profiles from waveform data.
The top left panel shows the velocity profile comparison between the true linear profile and the recovered profiles from automatic differentiation (AD) and the inverse PINN.
The top right panel shows the final wavefield reconstructed with each recovered profile.
The bottom panels show optimization histories for both methods.
Although the inverse PINN achieves reasonable velocity recovery (RMSE $3.18 \times 10^{-2}$), AD converges faster in this benchmark while avoiding a separate neural-network approximation to the wavefield.}
\label{fig:inverse}
\end{figure}

\Cref{fig:inverse} compares the two inverse approaches. The inverse PINN achieves reasonable velocity recovery (RMSE $3.18 \times 10^{-2}$), while AD converges faster and more reliably in this experiment. The AD approach requires roughly 80 lines of JAX code, compared with more than 500 lines for the PINN implementation and additional hyperparameter tuning.

PINNs may still help when measurements lie at irregular locations, when meshing the observation geometry is inconvenient, or when sparse noisy data must combine with residual constraints. For inverse problems where a reliable forward solver already exists, however, differentiating that solver provides a simpler baseline that we recommend trying before introducing a PINN.

\section{Operator Learning and the Data Requirement for DeepONet}
\label{sec:deeponet}

MLPs and PINNs approximate a single function $u(x)$ for a specified problem instance. Operator learning instead approximates a mapping between function spaces. Formally, it learns an operator $G \colon \mathcal{U} \to \mathcal{V}$; for a beam on elastic foundation, $G$ maps a load function $p(x)$ to the deflection function $w(x)$. Classical methods solve the governing equation for each new load function, while operator learning trains once on many $(p, w)$ pairs and then evaluates the learned operator for new loads.

Two leading architectures for this task are Deep Operator Networks (DeepONet) \citep{Lu2021DeepONet} and Fourier Neural Operators (FNO) \citep{Li2021FNO}. DeepONet uses a branch network to encode the input function $p(x)$ sampled at fixed sensor locations and a trunk network to encode the query locations where output is desired. The prediction is the inner product of the branch and trunk encodings, $\hat{w}(x) = \sum_{k=1}^{K} b_k(p) \, t_k(x)$, where $K$ is the latent dimension. Each output value $\hat{w}(x)$ therefore depends on the entire input function $p$ through the branch encoding. This global structure distinguishes operator learning from the pointwise residual construction used by PINNs.

The possible advantage of operator learning is amortization. The training cost is paid once, and the trained model is then queried many times. Whether this amortization is useful depends on the cost of the reference solver, the number of future queries, the cost of generating training data, and the inference cost of the trained network.

\subsection{Numerical Experiment on a Beam on Elastic Foundation}

A beam on elastic foundation (Winkler model) satisfies the Euler-Bernoulli equation with a linear spring restoring force.
\begin{equation}
EI w^{(4)} + k w = p(x)
\end{equation}
Here $w(x)$ is deflection, $p(x)$ is distributed load, $EI = 5 \times 10^6$ N$\cdot$m$^2$ is flexural rigidity, and $k = 10^5$ N/m$^2$ is the foundation modulus (modulus of subgrade reaction).
The beam is simply supported, so $w(0) = w(L) = w''(0) = w''(L) = 0$ with $L = 10$ m.

We generate a training set of 5000 random load profiles that combine Gaussian bumps, Fourier modes, and polynomials. For each load profile, we compute a finite difference solution of the fourth-order differential equation using 128 grid points and a five-point stencil for the fourth derivative.

\begin{figure}[h]
\centering
\includegraphics[width=0.85\textwidth]{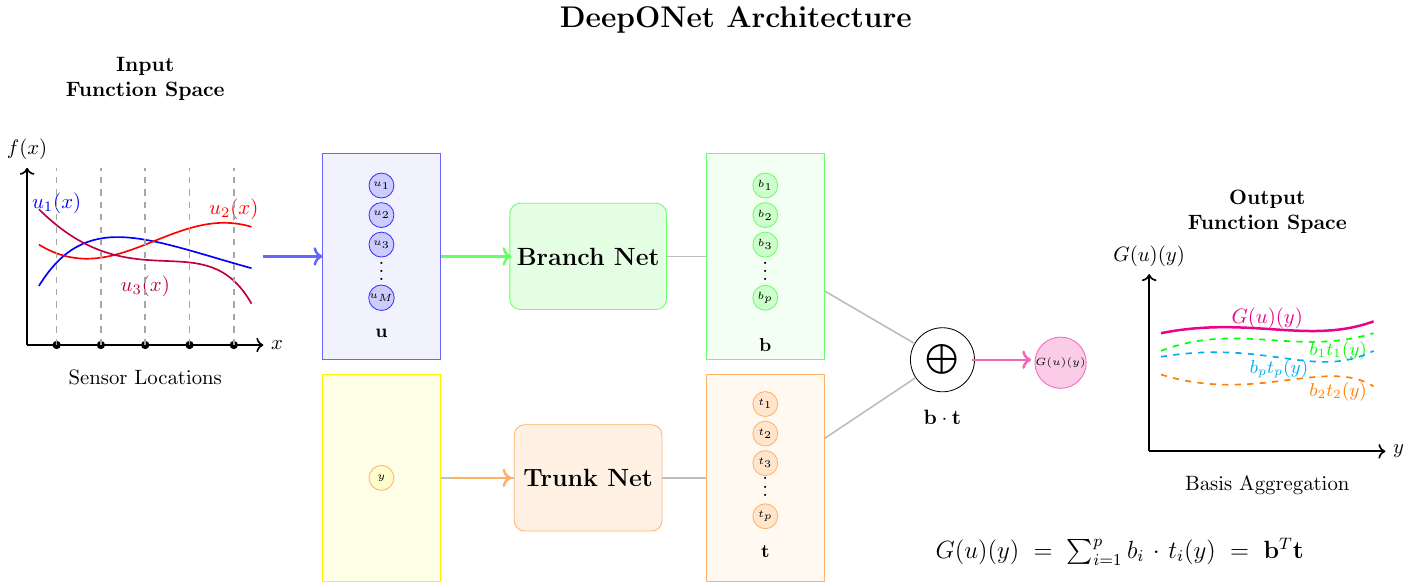}
\caption{Deep Operator Network (DeepONet) architecture.
The branch network encodes the input function (applied load $p(x)$) sampled at fixed sensor locations into a latent representation.
The trunk network encodes query locations where predictions are desired.
The two representations combine through a dot product to produce the output (deflection $w(x)$).
This architecture learns to map entire functions to functions, which enables fast prediction of beam deflections for arbitrary load patterns after offline training.}
\label{fig:deeponet_arch}
\end{figure}

\textbf{Architecture details.} The DeepONet consists of two parallel networks, a \textit{branch network} that encodes the input load $p(x)$ and a \textit{trunk network} that encodes query locations. The branch network takes the load sampled at 128 fixed sensor locations as input, processes it through 3 hidden layers of 256 neurons each with tanh activation, and outputs a 256-dimensional latent representation. The trunk network takes a single spatial coordinate $x$ as input, processes it through 3 hidden layers of 256 neurons, also with tanh activation, and outputs the corresponding 256-dimensional representation. The latent dimension is $K = 256$, and the dot product of the two representations gives the prediction (\cref{fig:deeponet_arch}). Training runs for 8000 iterations using the AdamW optimizer with cosine learning rate decay starting from $4 \times 10^{-4}$.

The FD solver completes each solution in 0.11 ms. DeepONet training requires 203 seconds, and inference takes 5.45 ms per query with normalized test MSE of $4.9 \times 10^{-3}$. In this benchmark, the trained DeepONet is about 50 times slower per query than the direct solver, so additional queries increase the total time gap after training has already been paid. For 10{,}000 load cases, FD total time is 1.1 seconds while DeepONet requires 258 seconds. The 203-second training overhead alone is equivalent to roughly 1.8 million FD solves at 0.11 ms each.

\subsection{When Operator Learning Makes Sense}

The break-even point depends on the relative cost of training, inference, and direct solution. On the Winkler beam tested here, DeepONet inference at 5.45 ms per query is already slower than FD at 0.11 ms, so amortization cannot recover the training cost. Operator learning becomes computationally attractive only when the underlying solver is expensive enough that the trained network is faster per query and when the same family of problems is queried many times. This condition is unlikely for geotechnical calculations that already run in microseconds or seconds, such as 1D consolidation, 2D seepage, or simple elastic settlement. It may arise in 3D soil-structure interaction models or digital-twin settings where repeated near-real-time predictions are required.

Even in the expensive-solver regime, operator learning should be compared with reduced-order and surrogate methods already used in mechanics, including proper orthogonal decomposition, kriging, and polynomial chaos expansion. The relevant comparison is the full workflow, including data generation, training, validation, and inference, rather than the trained neural operator alone against one full simulation.

\subsection{Limitations of DeepONet}

DeepONet has structural limitations beyond the cost comparison. Training requires thousands of input-output function pairs; the beam experiment used 5{,}000 load-deflection pairs, each from a separate forward solve. For the expensive forward models where operator learning is most attractive, this data-generation step may dominate the total cost before training begins.

The branch network samples the input function at fixed sensor locations.
After training, the network expects input functions sampled at exactly those locations.
A change in sensor placement, which happens often in field monitoring, requires retraining.
DeepONet also learns the operator for a specific domain geometry.
Changing the beam length, adding a notch, or modifying boundary locations invalidates the trained network.
Finite element solvers handle arbitrary geometries through mesh generation, whereas DeepONet embeds the geometry into its learned weights.

DeepONet also does not automatically encode symmetries such as translational invariance. A beam from $x = 0$ to $x = L$ and an equivalent beam from $x = 2$ to $x = 2 + L$ satisfy the same shifted governing equation under the same shifted loading. A standard DeepONet, however, treats the coordinate values as distinct inputs. Unless the architecture or training procedure accounts for that symmetry, the learning problem has greater effective complexity and requires more data to cover the same physical behavior.

These limitations relax when geometry, material properties, and sensor locations stay fixed. A bridge, building, or tunnel section instrumented for long-term monitoring has a fixed domain, while loading and boundary conditions vary through time. In that setting, a trained operator can serve as a fast surrogate once many queries offset the upfront training cost. \citet{Xu2024DeepONettunnel} demonstrated this type of application for real-time settlement prediction during tunnel construction using a multi-fidelity DeepONet that fuses simulation and monitoring data.

\section{Graph Network Simulators for Particle Dynamics}
\label{sec:gns}

Graph Network Simulators (GNS) take a local approach to learned simulation \citep{Sanchez2020GNS}, with related mesh-based methods including MeshGraphNets \citep{Pfaff2021MeshGraphNets}. Where DeepONet learns a global operator on whole functions, GNS learns a local update rule on a graph. Each node update depends on neighboring particles, and GNS applies the same learned rule throughout the graph. This locality is the source of the method's geometric transfer. A new geometry changes the graph connectivity, while the learned interaction rule remains the same.

The method discretizes the physical domain with nodes that represent material points and edges that connect neighboring particles.
The architecture follows an encode-process-decode structure.
The encoder embeds particle states (position, velocity, material properties) and boundary information into latent node and edge features.
The processor applies multiple rounds of message passing.
Each node aggregates information from its neighbors through learned functions, and this aggregation is where the network learns the physics of local interactions.
The decoder extracts accelerations from the processed node features.
A simple Euler integration step then updates positions and velocities for the next timestep.
Training uses trajectories from high-fidelity simulations (DEM, MPM) and minimizes the error between predicted and actual accelerations.

\begin{figure}[h]
\centering
\includegraphics[width=0.95\textwidth]{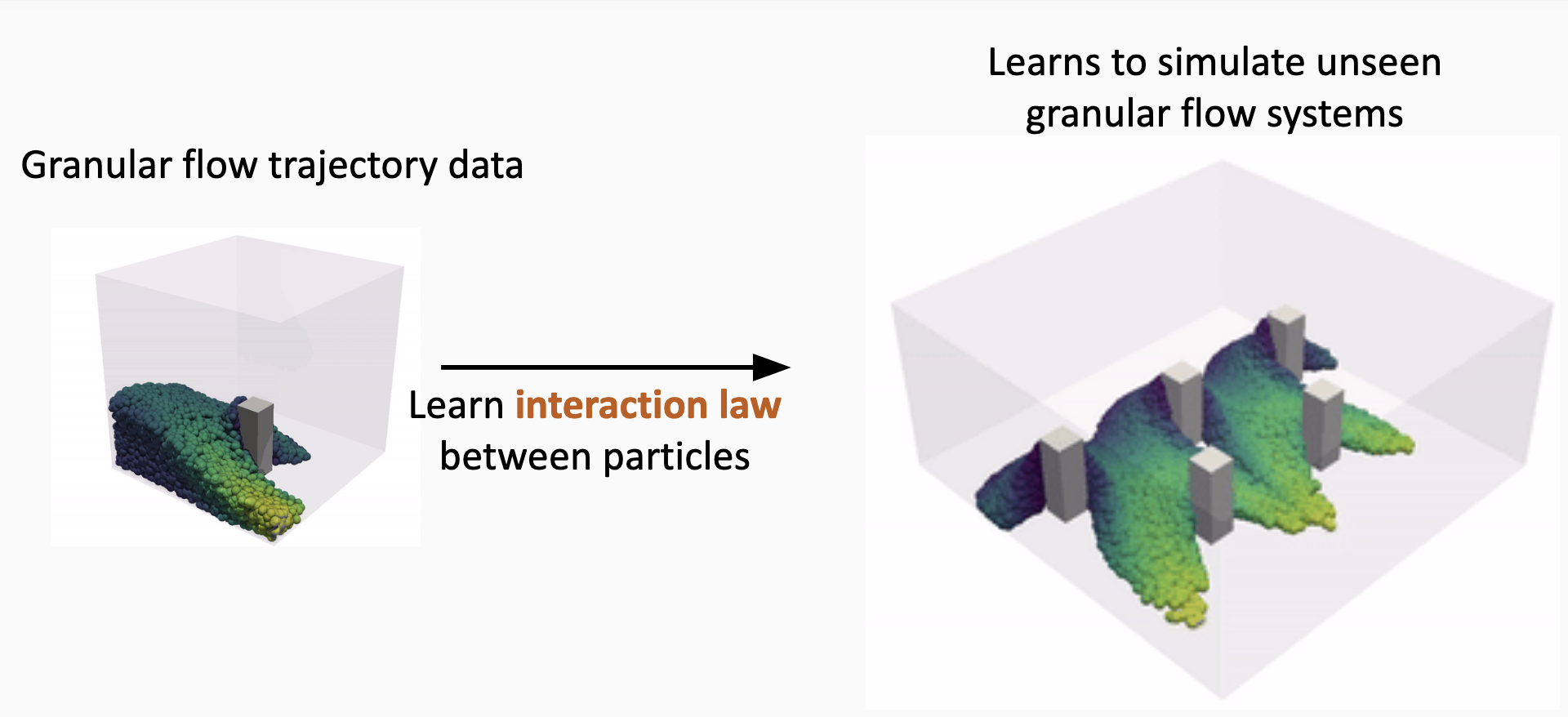}
\caption{GNS generalization to unseen boundary conditions.
The model trains on granular column collapse with a single barrier (top row) but accurately predicts flow dynamics with multiple barriers and different configurations not seen during training (bottom rows).
This geometry-agnostic behavior arises because GNS learns local interaction laws rather than global input-output mappings.
Adapted from \citet{Choi2024GNS}.}
\label{fig:gns_generalization}
\end{figure}

The graph structure can adapt to different particle configurations, as illustrated in \cref{fig:gns_generalization}. A GNS trained on granular column collapse with a single barrier configuration predicts flow dynamics with multiple barriers, different barrier positions, and barrier shapes not seen during training \citep{Choi2024GNS}. In this example, transfer arises from learned particle-particle and particle-boundary interactions instead of a fixed global input-output map.

\citet{Choi2024GNS} applied GNS to granular flows relevant to geotechnical engineering, demonstrating speedups of several thousand times over Material Point Method (MPM) simulations while maintaining accuracy within 5\% for column collapse problems.
They further developed differentiable GNS for inverse analysis, using automatic differentiation through the learned simulator to estimate material parameters from observed flow behavior \citep{Choi2024GNSInverse}.

Current GNS formulations have several limitations, starting with initialization. The simulator usually requires warmstarting because it uses a short history of particle positions or velocities to initialize predictions. That requirement precludes instantaneous evaluation from an arbitrary static configuration and complicates coupling with other simulation components.

Granular materials exhibit path-dependent behavior that GNS cannot capture.
Current stresses depend on the entire loading history rather than the current configuration alone.
GNS receives only the last few timesteps (typically 3 to 5 frames) as input.
A sample loaded in compression and then shear behaves differently from one loaded in shear alone, even when current configurations are identical.
A GNS trained on monotonic loading cannot generalize to cyclic loading or stress reversals \citep{Fransen2026ScientificML}.

Memory imposes another scaling limit. For $N$ particles with average degree $k$, GNS requires $O(Nk)$ memory per timestep, and training stores activations for backpropagation through time. A 1000-timestep simulation with $10^6$ particles can therefore exceed 100 GB of memory. Reported demonstrations commonly use $10^3$ to $10^4$ particles, which is two to three orders of magnitude smaller than the $10^6$ to $10^7$ particles often needed for foundation bearing capacity or shear band localization studies.

Distributed training adds another scaling constraint because multi-GPU GNS training requires domain decomposition of each trajectory, ghost nodes at partition boundaries, and management of particles that migrate between partitions. These spatial couplings make graph data more difficult to distribute than image batches. For geotechnical DEM problems with $10^4$ particles that already run in minutes, the training cost, data requirements, and approximation error of GNS are difficult to justify. The case for GNS is stronger when the reference solver takes hours, when engineers need to evaluate many related geometries, and when the local interaction rule transfers across those geometries.

\section{Discussion and Conclusion}
\label{sec:discussion}

The experiments point to a common requirement for SciML in geotechnical engineering. The right assessment depends on the engineering role the model will play. When the role is extrapolation, training accuracy is an inadequate metric. The MLP consolidation example shows that networks with low error over the observed interval can extrapolate according to activation-function tails rather than consolidation physics. The damped-oscillator PINN gives the same lesson for residual-based training. The residual constrains the learned function only at the sampled points, and it does not turn the MLP into a time-marching solver outside the collocation domain.

When the role is forward simulation, computational accounting must include data generation, training, inference, and retraining. On the 1D wave equation, the PINN is nearly 96{,}000 times slower than finite difference for a single forward solution and remains less accurate. DeepONet addresses the single-instance limitation by learning a global operator, although the beam benchmark shows that this benefit appears only when amortization is possible. If the direct solver is faster than the trained network at inference, additional queries cannot recover the training cost. GNS has a different advantage because its local graph update can transfer across particle configurations. Its warmstart requirement, path-history limitation, memory footprint, and distributed-training complexity remain important constraints for geotechnical-scale particle simulations.

The inverse-problem comparison gives a different outcome. When a reliable forward solver already exists, differentiating that solver preserves the discretized physics and avoids training a separate neural approximation to the state field. In the inverse wave benchmark, AD through the finite difference solver recovers the velocity profile in seconds with about 1\% relative error and less implementation complexity than the PINN. Reverse-mode AD still has costs, especially activation storage and the need for differentiable solver operations, but those costs come from the forward model rather than from an additional residual-training problem.

The practical implication is to select SciML methods by task. For pattern recognition within a validated data domain, such as CPT interpretation \citep{Hudson2023CPT}, anomaly detection, and correlation of sparse monitoring data, machine learning can be useful if validation respects site-level spatial correlation. For safety-critical prediction outside the observed domain, engineers should check the method against physics-based calculations, quantify uncertainty, and organize train-test splits by site or project rather than by randomly shuffled measurements. For forward analyses where traditional solvers already run in seconds, neural solvers need a demonstrated gain in transferability, uncertainty quantification, or integration with data before their training cost is justified. For inverse analyses, the natural baseline is a differentiable implementation of the existing physics-based solver.

Hybrid approaches remain important in this assessment \citep{Fransen2026ScientificML}. Neural networks can learn constitutive response from element tests and embed that response in finite element solvers that enforce equilibrium. Learned representations can compress random fields for stochastic analysis. Differentiable solvers can assimilate sparse sensor data while retaining the governing discretization. These uses align with geotechnical engineering practice more directly than replacing mature solvers with neural approximations solely because the trained model can be evaluated quickly. The benchmark results here therefore support a bounded conclusion. SciML has value in geotechnical engineering when its data domain, physical constraints, computational cost, and validation design are explicit, while neural-network solvers require evidence before they replace established numerical methods.

\section*{Data Availability Statement}

All code, numerical experiments, and Jupyter notebooks used in this study are publicly available at \url{https://github.com/geoelements/ml-geo}.

\section*{Acknowledgments}

The author thanks Prof. Ellen Rathje (UT Austin) for insightful discussions on machine learning applications in earthquake geotechnics and spatial correlation in geotechnical data.

\bibliographystyle{apalike}
\bibliography{references}

\end{document}